\documentclass[5p,twocolumn]{elsarticle}
\usepackage{graphicx,latexsym}
\usepackage{dcolumn}
\usepackage{amssymb,amsmath,bm}
\usepackage{subfigure}
\usepackage{braket}
\usepackage{siunitx}
\usepackage{array}
\usepackage{float}
\usepackage[nopar]{lipsum}
\usepackage{caption}
\usepackage[super]{nth}

\newcommand{\angstrom}{\text{\normalfont\AA}}
\usepackage{hyperref}
\hypersetup{
    pdfnewwindow=true,       
    colorlinks=true,         
    linkcolor=blue,          
    citecolor=blue,          
    filecolor=magenta,       
    urlcolor=black           
}

\usepackage[normalem]{ulem}

\def\sec#1{Sec.\ \ref{#1}}

\def\fig#1{Fig.\ \ref{#1}}

\journal{}

\begin{document}

\begin{frontmatter}


\title{Modeling electronic, mechanical, optical and thermal properties of
	graphene-like BC$_6$N materials: Role of prominent BN-bonds}

\author[a1,a2]{Nzar Rauf Abdullah}
\ead{nzar.r.abdullah@gmail.com}
\address[a1]{Division of Computational Nanoscience, Physics Department, College of Science, 
             University of Sulaimani, Sulaimani 46001, Kurdistan Region, Iraq}
\address[a2]{Computer Engineering Department, College of Engineering, Komar University of Science and Technology, Sulaimani 46001, Kurdistan Region, Iraq}

\author[a1]{Hunar Omar Rashid}

\author[a4]{Chi-Shung Tang}
\address[a4]{Department of Mechanical Engineering,
	National United University, 1, Lienda, Miaoli 36003, Taiwan}

\author[a5]{Andrei Manolescu}
\address[a5]{Reykjavik University, School of Science and Engineering,
	Menntavegur 1, IS-101 Reykjavik, Iceland}

\author[a6]{Vidar Gudmundsson}   
\address[a6]{Science Institute, University of Iceland,
	Dunhaga 3, IS-107 Reykjavik, Iceland}

%

\begin{abstract}

We model monolayer graphene-like materials with BC$_6$N stoichiometry where the bonding between the B 
and the N atoms plays an important role for their physical and chemical properties.  
Two types of BC$_6$N are found based on the BN bonds: In the presence of BN bonds, an even number 
of $\pi$-bonds emerges indicating an aromatic structure and a large direct bandgap appears, while 
in the absence of BN bonds, an anti-aromatic structure with an odd-number of $\pi$-bonds is found 
resulting a direct small bandgap. The stress-strain curves shows high elastic moduli and tensile strength 
of the structures with BN-bonds, compared to structures without BN-bonds. 
Self-consistent field calculations demonstrate that BC$_6$N with BN-bonds is energetically more stable 
than structures without BN-bonds due to a strong binding energy between the B and the N atoms, while 
their phonon dispersion displays that BC$_6$N without BN-bonds has more dynamical stability. 
Furthermore, all the BC$_6$N structures considered show a large absorption of electromagnetic radiation 
with polarization parallel to the monolayers in the visible range. Finer detail of the absorption 
depend on the actual structures of the layers. 
A higher electronic thermal conductivity and specific heat are seen in BC$_6$N systems caused by hot 
carrier--assisted charge transport. This opens up a possible optimization for bolometric applications 
of graphene based material devices.
\end{abstract}

\begin{keyword}
Energy harvesting \sep Thermal transport \sep Graphene \sep Density Functional Theory \sep Electronic structure \sep  Optical properties \sep  and Stress-strain curve 
\end{keyword}

\end{frontmatter}

\section{Introduction}

Material structures at the nanoscale with a combination of carbon atoms, C, with its
neighboring Boron, B, and Nitrogen, N, atoms in the periodic table forming BCN \cite{Rubio2010}, 
BC$_3$N \cite{doi:10.1142/q0078}, and BC$_6$N \cite{MORTAZAVI2019733} have shown interesting physical properties and lead to numerous potential 
technological applications in the fields of lithium ion batteries \cite{doi:10.1021/nn101926g}, electrocatalysis \cite{PMID:22431416}, sensors \cite{doi:10.1021/acsnano.6b08136}, nano-electronics devices, optical devices \cite{KUMAR2010152}, and field emission.

Experiments and calculations found that the physical properties of B$_x$C$_y$N$_z$ depend on the 
concentration of B, C, and N atoms in the lattice \cite{PhysRevB.51.11229, PhysRevB.49.5081}.
For instance, the stress-strain characteristics of BC$_6$N have shown
anisotropy in the mechanical response when uniaxial
tensile simulations were conducted along the armchair and the zigzag directions \cite{MORTAZAVI2019733}. 
In addition, increasing the thickness of BC$_6$N results in a decreased bandgap, and the application of a 
perpendicular electric field decreases the bandgap even more. Semiconductor to insulator transition is thus found \cite{BAFEKRY2020113850}. The bandgap opening can be tuned by the concentration ratio of B and N atoms, 
in which the variation of the bandgap with the B and N dopant concentration is almost linear \cite{Rani2014}.

It has been found that the inversion symmetry breaking of BC$_6$N with high carrier mobility leads 
to a pair of inequivalent valleys with opposite Berry curvatures in the vicinities of vertices \cite{C8NR03080D}. This leads to a direct bandgap with $1.833$~eV, which is beneficial for realization 
of a valley Hall effect~\cite{C8NR03080D}.

One of the important optical characteristics of graphene is its capability to function as an absorbent 
for a wide range of electromagnetic radiation \cite{Marra2018}. The optical study of graphene and related
materials is important for understanding their electronic structures and excitation spectra.
Consequently, there are many potential applications of optical and photonic effects in which 
graphene has been used \cite{Loh2010, doi:10.1063/1.5051796, ABDULLAH2020113996, ABDULLAH2019102686}.
The sensitivity of the electronic properties of graphene to electromagnetic adsorption indicates a very good candidate for photonic application such as sensors \cite{Schedin2007}, and
the low electron energy loss spectroscopy (EELS) allows
graphene to be used for detecting changes in the electronic structure 
with a high spatial resolution \cite{PhysRevB.92.125147}.
BC$_6$N also presents highly promising optical characteristics while
individual B or N doping of graphene does not significantly influence the absorption spectra.
However, significant red shift in absorption towards the visible range of the radiation at high 
BN doping is found to occur \cite{RANI201428}.

\lipsum[0]
\begin{figure*}[]
	\centering
	\includegraphics[width=0.9\textwidth]{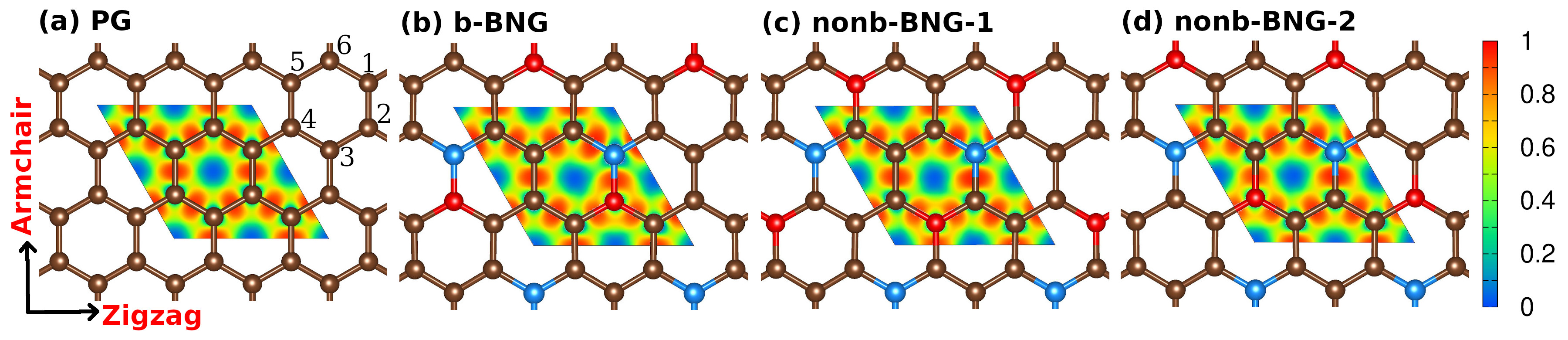}
	\caption{Pure graphene (PG) nanosheet (a), and b-BNG (b), nonb-BNG-1 (c) and nonb-BNG-2 (d) 
		nanosheets were the C atoms are brown, B atoms are blue, and N atoms are red. The contour plots demonstrate the electron localization function (ELF). The numbers from 1 to 6 display the position of atoms in the hexagon. The positions 1, 2, and 3 in the hexagon are called ortho-, meta-, and para-position, respectively.}
	\label{fig01}
\end{figure*}

Thermoelectric measurements of BC$_6$N show that the material behaves
as a p-type or an n-type semiconductor with a high Seebeck coefficient of 300~$\mu$V/K
at room temperature \cite{doi:10.1021/cm950471y}. It has also been revealed that the B$_x$C$_y$N$_z$
system marks up to a 20 times larger Seebeck coefficient than pure graphene \cite{doi:10.1063/1.4820820}.

To alter the properties of pure graphene, we investigate a substitutional BN codoping, due to the radii 
similarity between them and C, along with the resultant BC$_6$N semiconductor. 
The main idea in this work is the effect of the bonding between the B and N dopants in the BC$_6$N on 
its physical characteristics. We analyze the electronic, the mechanical, the optical, 
and the thermal characteristics, depending on the composition and the atomic arrangement of the B, C, and N 
atoms in the lattice. 
In addition, we investigate the stability of the resulting structures, 
and how the internal interaction energies influence their properties. 
We observe to which extent the BN co-doping positions within the graphene sheets modify 
all their physical characteristics.

In \sec{Sec:Model} the structure of BC$_6$N nanosheet is briefly over-viewed. In \sec{Sec:Results} the main achieved results are analyzed. In \sec{Sec:Conclusion} the conclusion of the results is presented.

\section{Model and Computational Details}\label{Sec:Model}

We perform first-principle calculations on monolayer BC$_6$N nanosheets using
the Perdew-Zunger (PZ) exchange–correlation functionals \cite{PhysRevB.23.5048}.
Norm-conserving pseudopotential is used in the current work \cite{PhysRevLett.43.1494} with 
a kinetic energy cutoff \SI{10}{\electronvolt}.
The atomic positions are fully optimized by implementing a quasi-Newton
algorithm for atomic force relaxation. The structure is relaxed until the total force on each atom is less than \SI{0.02}{eV\per{\angstrom}} and the mesh grid of the atomic force relaxation is  $12\times12\times1$. All calculations are performed using the Kohn–Sham version of the Density Functional Theory (DFT) \cite{PhysRev.140.A1133} implemented in the Quantum Espresso (QE) package \cite{giannozzi2009quantum,giannozzi2017advanced}.
We consider a large vacuum interval, 20~$\angstrom$, between the periodic slabs of BC$_6$N to avoid 
interaction between adjacent layers. 

For each structure, the integration over the Brillouin zone was calculated by QE on an $8\times8\times1$ grid, using Bloch's theorem for the Self Consistent Field (SCF) calculation. Then a denser mesh-grid of $100\times100\times1$ was used for the density of states (DOS) calculations, and the band structure was calculated along a smooth path along ($\Gamma$-K-M-$\Gamma$). 
In addition, XCrySDen, a crystalline and molecular structure visualization program, is used for 
visualization of the results of our models \cite{KOKALJ1999176}.

In order to explore the optical properties of the system, we calculate the dielectric tensor with QE 
using the energy interval from 0 to \SI{20}{\electronvolt}. The imaginary part uses the response function 
derived from a perturbation theory with adiabatic tuning, while the real part is evaluated using the 
Kramers-Kroning transformation \cite{deL.Kronig:26}.

Consequently, we analyze the post-processed results from the QE with the Boltzmann transport properties (BolzTraP) package which calculates the semi-classic transport coefficients \cite{madsen2006boltztrap}. 
We can thus calculate the electronic thermal conductivity, electric conductivity, and specific heat 
\cite{RASHID2019102625, ABDULLAH2020126350}.

\section{Results}\label{Sec:Results} 

In this section, our main results for pristine and BN-codoped graphene forming BC$_6$N are shown. After modeling a pure graphene (PG) structure, we constructed a BN-codoping within the graphene 
to make three configurations of BC$_6$N: B is fixed at the ortho position (blue color) and N is put at 
the meta, or the para position (red color) with respect to B, as shown in \fig{fig01}. 
In the first doped structure, it is assumed that there is bonding between B and N atoms (BN-bond) which is identified as b-BNG here (\fig{fig01}(b)). 
Furthermore, assuming there is no BN-bond in the structure, we consider one (two) carbon atom(s) located between the B and the N atom, identified as nonb-BNG-1 and nonb-BNG-2, respectively. The number 
1 and 2 here refers to the number of carbon atom(s) between the B and the N atom in the clockwise direction.
The relative positions of the B and N around all the C hexagon is a key factor for the physical properties of the doped graphene. 
As we know, each carbon atom has 4 electrons in the outer-shell (out of 6), 3 are bonded with neighboring atoms ($\sigma$-electrons) and the \nth{4} electron (a delocalized $\pi$-electron) which is above and below the honeycomb, contributes to conduction, and the $\pi$-orbitals overlap, strengthening the bonding.  The highly mobile $\pi$-electrons act as mass-less fermions that satisfies the Dirac equation.

\lipsum[0]
\begin{figure*}[h]
	\centering
	\includegraphics[width=0.9\textwidth,height=4cm]{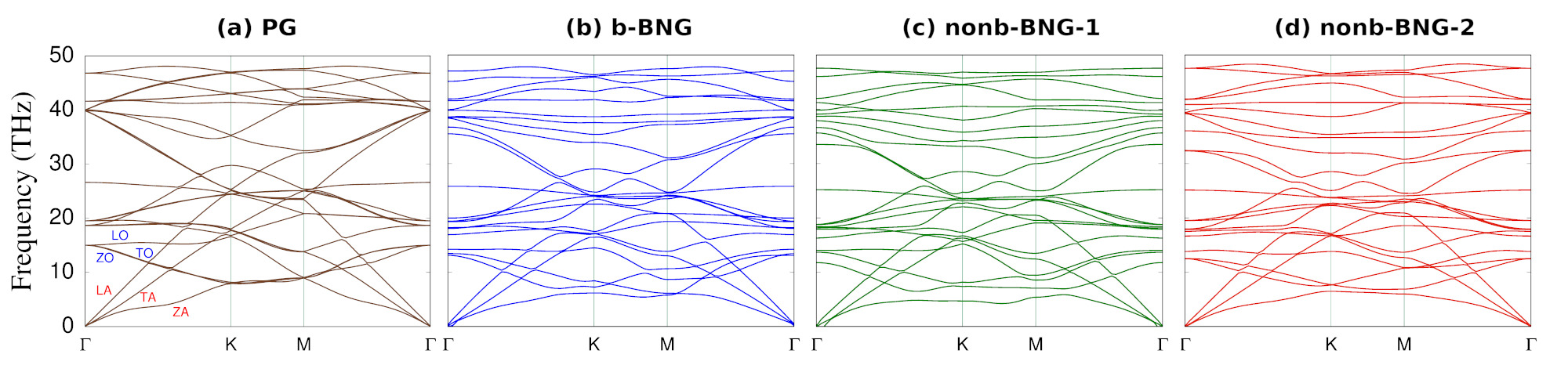}
	\caption{The calculated phonon band structure of PG (a), b-BNG (b), nonb-BNG-1 (c), and nonb-BNG-2 (d) structures along high symmetry of first Brillouin zone.}
	\label{fig02}
\end{figure*}

The lattice constants of b-BNG, nonb-BNG-1, and nonb-BNG-2 nanosheets are approximately equal to 
$4.855$, $4.851$, and $4.853$~${\angstrom}$, respectively. 
In the case of b-BNG, the average lengths of the C-C, the C-B and the C-N bonds are found to be $1.393$, $1.442$ and $1.378$~${\angstrom}$, respectively. 
For nonb-BNG-1, the average bond length of C-C, C-B and C-N bond lengths are $1.395$, $1.430$ and $1.389$~${\angstrom}$, respectively. In the last doped structure, nonb-BNG-2, the average bond lengths of C-C, C-B and C-N bond are $1.381$, $1.4302$ and $1.411$~${\angstrom}$, respectively. The bond lengths between the C, N and B atoms will help to understand better the mechanical properties of the BC$_6$N nanosheets as will be shown later. The number of C-C, C-B, C-N, and B-N bonds in a nanosheet will also influence the stability of the structure.

In order to understand better the bonding between the C, B, and N atoms, we provide the electron localization function (ELF) in \fig{fig01} (contour plot) \cite{Edgecombe1990}. A high ELF on a particular bond indicates a greater stiffness or rigidity of that bond. The ELF is high for the C-C bonds in PG. In addition, it is higher for a C-N bond compared to a C-B bond in a BN-codoped graphene structures \cite{GAO2018194}.

\subsection{Structural stability}

The binding energies of C-B and C-N are smaller than those of C-C and B-N. The more stable structure will thus have a lesser number of C-B and C-N bonds in the nanosheets \cite{C2NR11728B}. Therefore, the b-BNG is a relatively more stable structure than the nonb-BNG because of fewer C-B and C-N bonds in b-BNG. This observation shows the B–N bond is more stable than the B–C or the N–C bond. 
Total energy calculations are utilized to study the structure stability.
Our DFT calculations show that the formation energy of the b-BNG structure is $-191.721$~eV which is relatively smaller than that of 
the nonb-BNG-1, $-190.533$~eV, and the nonb-BNG-2, $-190.644$~eV. This demonstrates that b-BNG is  energetically more stable than a nonb-BNG structure. 

\subsection{Interaction energy}

The interaction energy of each considered structure can be obtained from the total energy \cite{RASHID2019102625}. 
The SCF calculations of our systems show that the b-BNG structure has a higher attractive interaction comparing to the nonb-BNG structures. This would be expected because the distance between the B and N atoms in b-BNG, the bonding length of B-N with $1.411$~${\angstrom}$, is small compared to the distance between the B and N atoms in nonb-BNG-1 with $2.432$~${\angstrom}$, and nonb-BNG-2 with $2.802$~${\angstrom}$.

\subsection{Dynamic stability}

The dynamic stability of BC$_6$N nanosheets can be checked by calculating the phonon dispersion relations presented in \fig{fig02} for the PG (a), b-BNG (b), nonb-BNG-1 (c) and 
nonb-BNG-2 (d). In general, the phonon dispersion of a unit-cell of graphene containing two atoms at A and B positions consists of six modes: Three acoustic and three optical modes. Our $2\times2\times1$ graphene supercell consists of 8 atoms leading to $24$ modes in the phonon band structure.

The acoustic modes of PG represented by ZA, TA, and LA (the lowest three modes) are isotropic for the long waves ($q \rightarrow \Gamma$) which is required for the symmetry characteristics of pure graphene \cite{FALKOVSKY20085189}. The next three modes represented by 
ZO, TO and LO are optical modes. In PG, the LO/TO at $\Gamma$, LO/LA at $K$ and 
ZO/ZA at $K$ are double degenerate indicating the symmetry of the unit cell.  In the three forms of the BC$_6$N structures here, we found that these modes are split, which demonstrates the symmetry breaking of the crystal. 

\lipsum[0]
\begin{figure*}[h]
	\centering
	\includegraphics[width=0.65\textwidth]{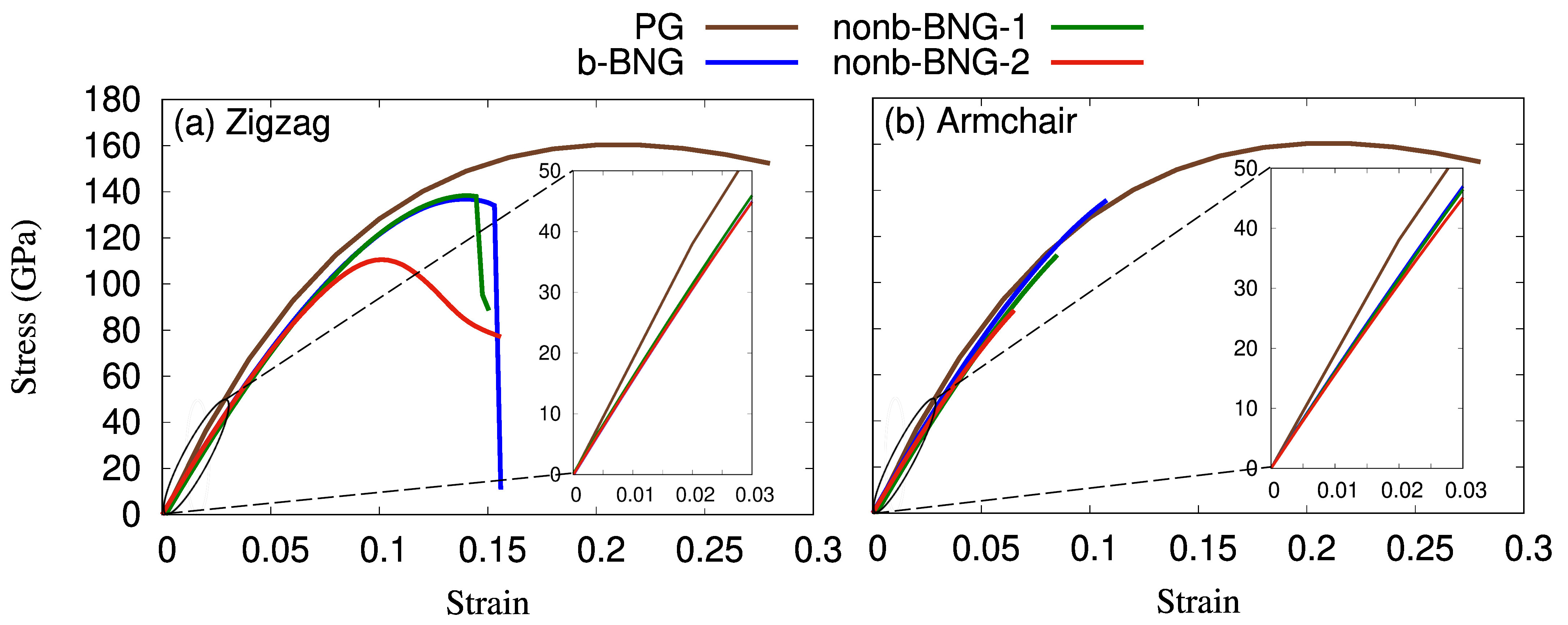}
	\caption{Stress-strain curves for PG (brown), b-BNG (blue), nonb-BNG-1 (green), and nonb-BNG-2 (red) in the zigzag (a) and the armchair (b) direction. The inset shows linear and elastic region of stress-strain curve.}
	\label{fig03}
\end{figure*}

The positive values of the phonon band structure of PG and nonb-BNG2-2 indicate the more dynamically stable structures of the nanosheets, while very small negativity in ZA mode of the b-BNG and the nonb-BNG-1 show relatively less dynamically stable structures. The ZA is the main mode responsible for the lattice thermal conductivity in the system \cite{Zhang2372}.

\subsection{Mechanical properties}

To calculate the mechanical behavior, the fully relaxed
simulation cell is exposed to a uniaxial tensile load of a constant strain rate. 
The stress-strain curves for three BC$_6$N structures are plotted together with 
the one for the PG in \fig{fig03} for the $x$-, the zigzag, (a) and the $y$-, 
the armchair direction (b).

\lipsum[0]
\begin{figure*}[h]
	\centering
	\includegraphics[width=0.9\textwidth,height=4cm]{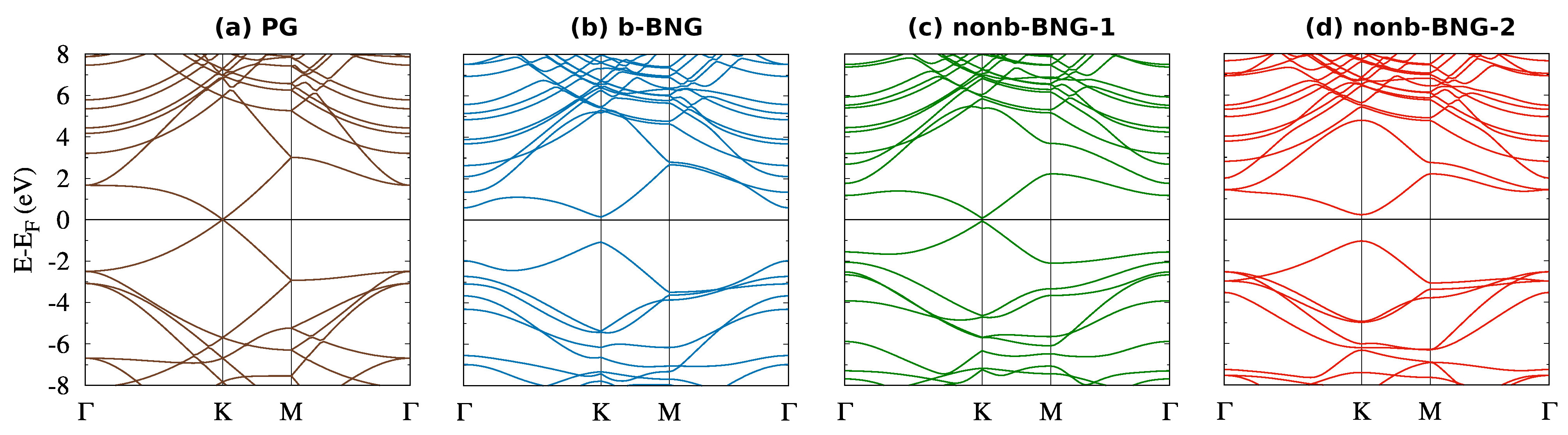}
	\caption{Band structures of PG (a), b-BNG (b), nonb-BNG-1 (c), and nonb-BNG-2 (d) with 0, $1.209$, $0.133$, $1.273$~eV bandgap, respectively. The Fermi energy is set to zero.}
	\label{fig04}
\end{figure*}

It can be clearly seen that the linear elastic regime for pure graphene (brown) ends at $\approx 2\%$ 
(see inset of \fig{fig03}), where the stress-strain relation becomes nonlinear for both load directions. 
In the all BC$_6$N structures, the stress is linearly proportional to strain up to $\approx 3\%$, and becomes nonlinear beyond $3\%$ for both the zigzag and the armchair directions.
In the linear regime the samples have elastic characteristics and 
both the pure and the BC$_6$N structures undergo only elastic deformations.
We notice that the pure graphene sample has a much higher failure strain of $28\%$ with 
stress $\approx 152$~GPa in both directions. The failure strain of PG is in agreement
with other stress-strain curve calculations done by molecular dynamics simulations \cite{Mortazavi_2014} and experimental results~\cite{Lee385}. 
The falling parts of blue, green and red lines, indicate a generation of 
cracks forming in the doped structures. The initial softening of the red curve 
indicates a higher ability of stretching in the structure before the fracture point
is reached \cite{Grantab946}.

The BC$_6$N structures all fail between $14\%$
and $17\%$ strain with stress between $76$ to
$138$~GPa in the zigzag direction, while in the armchair direction they 
fail between $6.5\%$ and $10\%$ with an ultimate stress $88$ and $136$~GPa. 
This indicates that b-BNG has larger internal stress than non-BNG due to the BN-bonds in the armchair direction and relatively high stress in the zigzag direction.   

We notice that the stress-strain curves of the three BC$_6$N samples are almost identical
at the failure strain in the zigzag direction, and the ultimate strength among the three
samples differs by less than $5\%$. Furthermore, the three BC$_6$N samples are different at the failure strain in the armchair direction. This is a convincing evidence that our sample size is sufficiently large to be statistically representative.

\subsection{Electronic Properties}

In this section, we focus on the electronic band structure. 
The length of the B-N bond is slightly greater than the C-C bond length leading to a good doped system with minimal internal stress, but the presence of the B-N together with the C-B and the C-N bonds in the graphene structure breaks the inversion symmetry leading to the opening of a band gap. At the same time, the bonding between the C, B and N atoms in the graphene structure plays an important role in the electronic characteristics. It means that
the relative positions of the B and N around all the C hexagon influence the bandgap as is 
shown in \fig{fig04}, where the band structure of the PG (a), b-BNG (b), nonb-BNG-1 (c), and nonb-BNG-2 (d) are presented.
As expected, the band structure of the PG has a vanishing bandgap between the valence and conduction bands at the K-point. It thus shows the semiconductor behavior of PG with 
a vanishing gap.

It can be seen from \fig{fig04} that the bandgap of b-BNG and nonb-BNG-2 is larger than that of nonb-BNG-1. This can be explained by the aromaticity which introduces a charge delocatization in the BC$_6$N structures. In b-BNG and nonb-BNG-2, an even number of $\pi$-bonds are found which 
indicates aromatic system while nonb-BNG-1 is anti-aromatic structure with an odd-number of $\pi$-bonds.
Interestingly, band gap tuning has been attained for graphene systems with adsorbed higher
aromatic molecules \cite{doi:10.1021/jp302293p}. 

\lipsum[0]
\begin{figure*}[htb]
	\centering
	\includegraphics[width=0.4\textwidth]{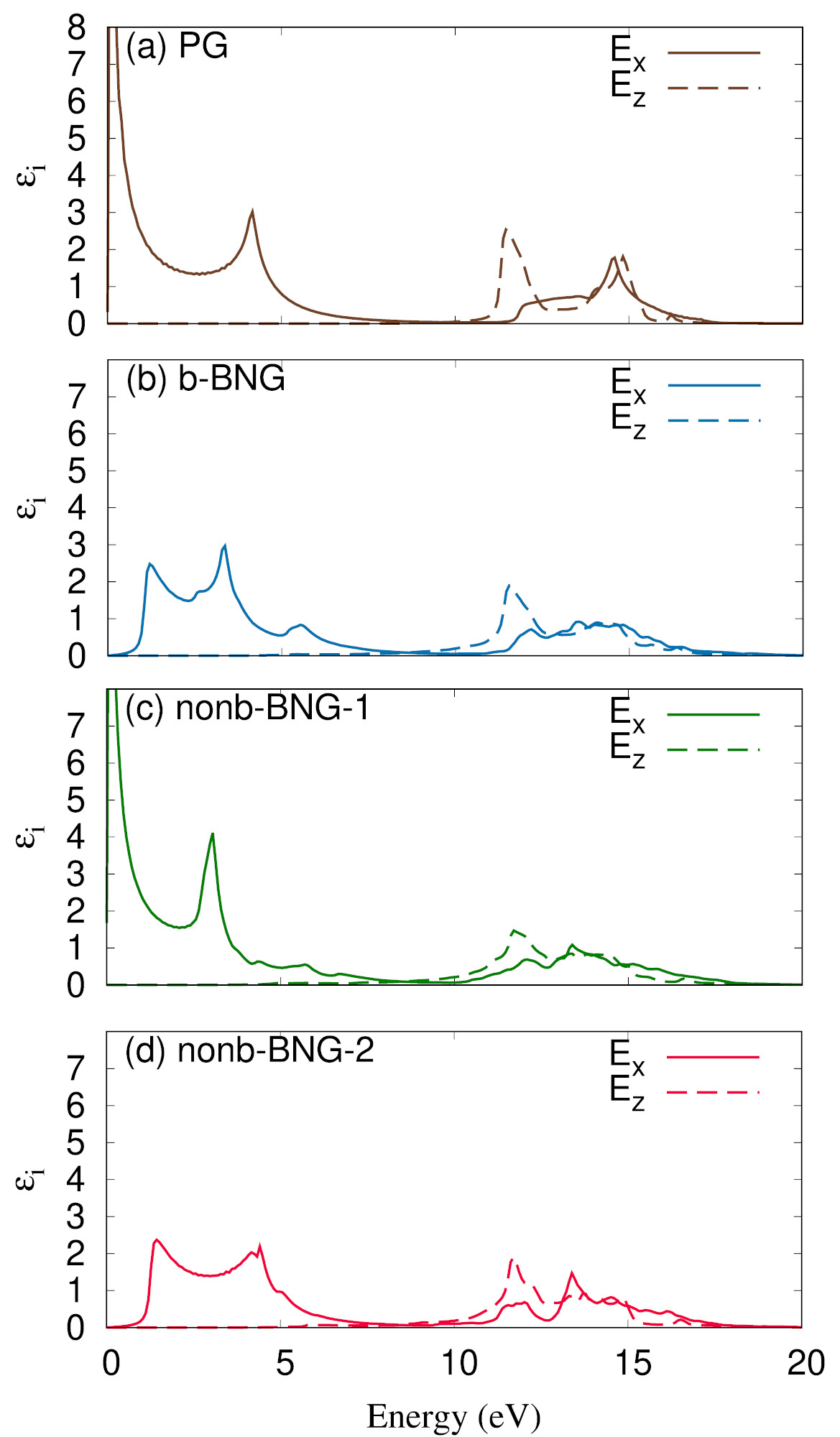}
	\includegraphics[width=0.29\textwidth]{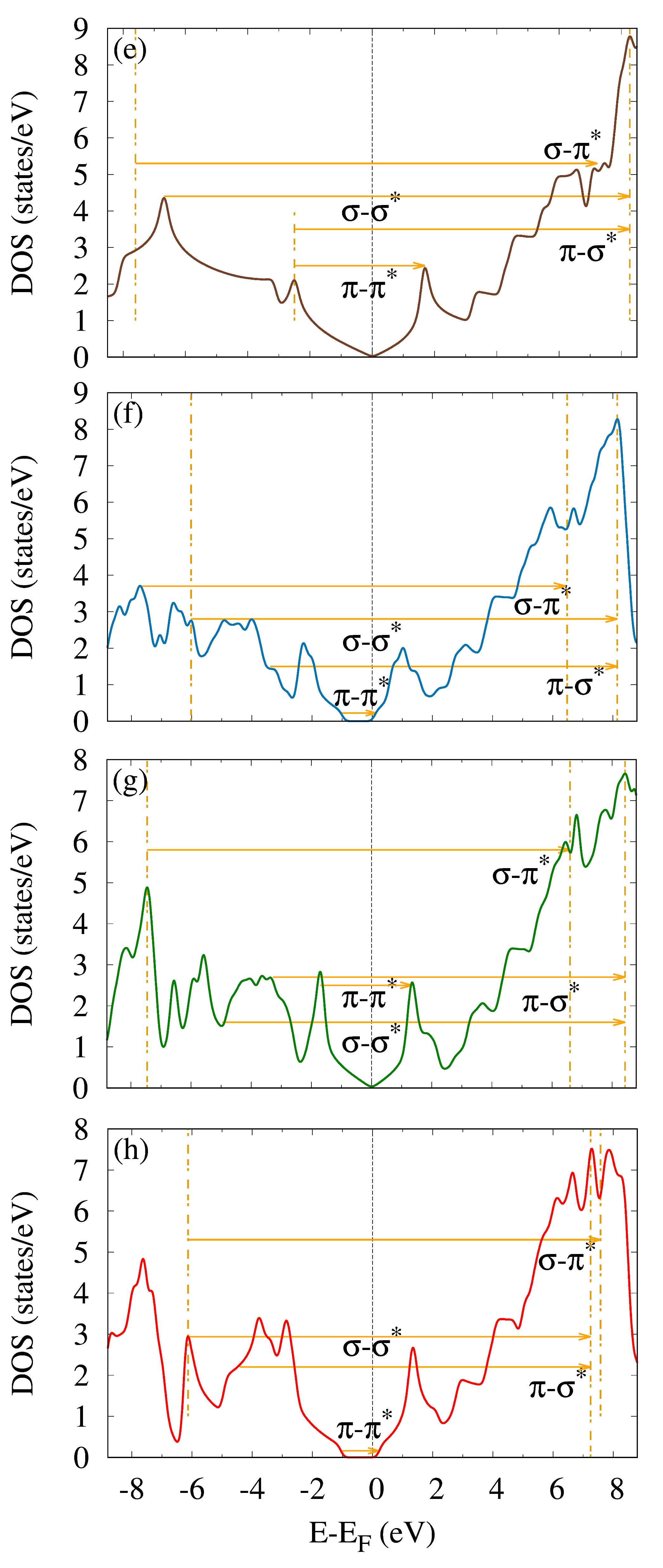}
	\caption{Imaginary part of dielectric function of PG (a), b-BNG (b), nonb-BNG-1 (c), and nonb-BNG-2 (d). Density of states for PG (e), b-BNG (f), nonb-BNG-1 (g), and nonb-BNG-2 (h).}
	\label{fig05}
\end{figure*}

Another method for opening a bandgap is the application of the Diels-Alder reaction (DA reaction) which introduces a covalent modification of graphene.
In the b-BNG and nonb-BNG-2 where the B and N atoms are substitutionally doped at the 1,2 or 1,4 positions of the hexagon, the DA reaction leads to the simultaneous transformation of a pair
of sp$^2$ carbons in the honeycomb lattice to either 1,2- or 1,4-sp$^3$
carbon centers, an AD reaction. Therefore, a structure with the $\pi$-electrons at neighboring C-atoms confined to local C-C $\pi$-bonds in the b-BNG and nonb-BNG-2 structures are found \cite{doi:10.1021/jp2016616}.
This opens up a bandgap \cite{SARKAR2012276}.
In addition, the strong polarity of the BN bonds in b-BNG may widen the bandgap.
In the nonb-BNG-1 structure, the B and N atoms located at position 1 and 3 of the
hexagon, imply two all-C segments, each with an odd number of $\pi$-electrons, in the hexagons leading to a very small bandgap \cite{doi:10.1021/jp2016616, Bhandary2012GrapheneBoronNC, C6CP03723B}.
These transformations of the covalent bonds are called sublattice symmetry breaking that opens up the bandgap \cite{Dollfus_2015}. Our results indicate that both the valence band maxima (VBM) and and conduction band minima (CBM) of b-BNG, nonb-BNG-1 and nonb-BNG-2 occur at the K-point, resulting a direct bandgap with $1.209$, $0.133$, $1.273$~eV, respectively.

\subsection{Optical Properties}

It is well known that the optical properties of graphene do vary with the number of layers.
For instance, mono-layer graphene can absorb light over a wide range of frequencies.
The imaginary part of the dielectric function, $\epsilon_i$, representing the absorption spectra is presented in \fig{fig05}(a-d) for the aforementioned structures shown in \fig{fig01}, with 
the electric field parallel, E$_x$, (solid line) or perpendicular, E$_z$, (dashed line) to the plain structure.
In order to explain the origin of the peaks in $\epsilon_i$, as contributions of specific transitions between bands, we present in \fig{fig05}(e-h) the density of states for the corresponding four structures. The location of the most important transitions is approximately indicated in the figures of DOS by golden arrows.

In the case of the PG (brown color), two expected peaks in the complex dielectric function for 
parallel electric field are observed (solid brown line) at $4.16$ and $14.59$~eV corresponding to the $\pi \rightarrow \pi^*$ and $\sigma \rightarrow \sigma^*$ transitions, respectively. The first peak has been confirmed experimentally at $4.5$~eV using spectroscopic ellipsometry analysis \cite{PhysRevB.81.155413}. The slight shift in the first peak in our result can be attributed to the neglect of the interaction between the graphene film and the substrate. In addition, the position of the second peak at $14.59$~eV in our calculation is very close to the experimental result at $14.6$~eV for free-standing graphene films~\cite{PhysRevB.77.233406}. 
\lipsum[0]
\begin{figure*}[htb]
	\centering
	\includegraphics[width=0.9\textwidth]{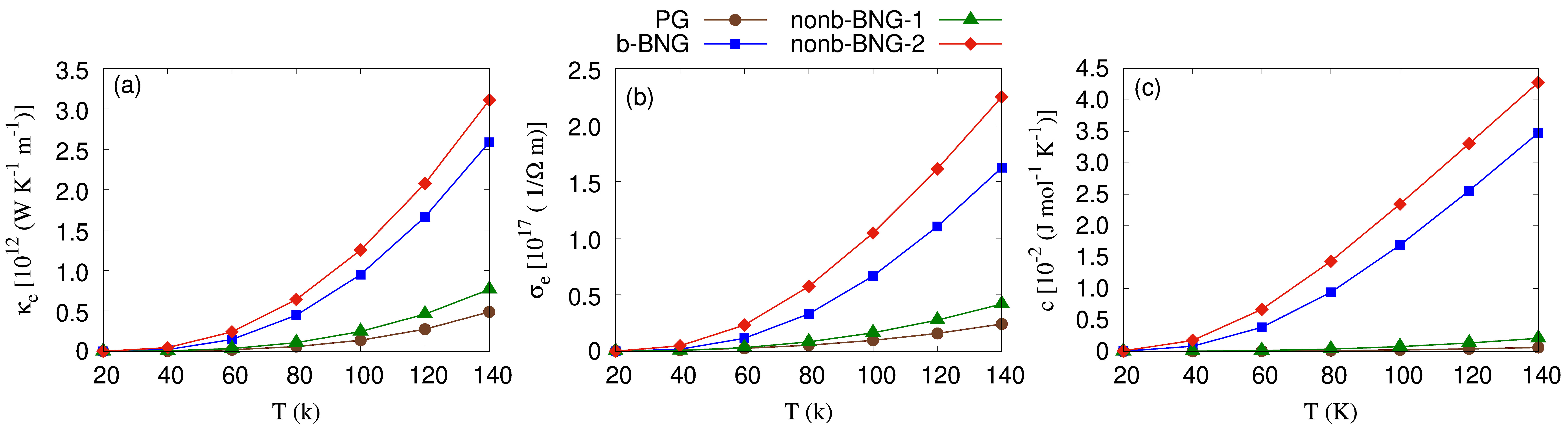}
	\caption{Electronic thermal conductivity (a), electronic conductivity (b), and specific heat (c) are plotted for PG (brown), b-BNG (blue), nonb-BNG-1 (green), and nonb-BNG-2 (red) near charge neutrality point.}
	\label{fig06}
\end{figure*}
The intensity ratio of these two peaks is $\epsilon_{i,L}/\epsilon_{i,R} = 3.015/1.778 = 1.695$ as is expected for monolayer graphene. The intensity ratio indicates that the $\pi \rightarrow \pi^*$ transition is stronger than the $\sigma \rightarrow \sigma^*$ transition.  
In the presence of a parallel electric field (brown dashed line) two peaks at 
$11.46$ and $14.83$~eV are found corresponding to the $\pi \rightarrow \sigma^*$ and the $\sigma \rightarrow \pi^*$ transitions, respectively.

In the BC$_6$N structures the rates of the transitions are changed due to the modification of 
the band structures. In the case of b-BNG and nonb-BNG-2, when there is a large bandgap, 
the $\pi\rightarrow\pi^*$ peak is shifted to $1.203$ and $1.44$~eV in the case of parallel electric field, respectively. The second peak is slightly smeared out and located at $3.36$ and $4.41$~eV for 
b-BNG and nonb-BNG-2, respectively, indicating a weaker $\sigma \rightarrow \sigma^*$ transition.
We note that the first peak in b-BNG and nonb-BNG-2 is located in the visible region. 
It demonstrates that these BC$_6$N monolayers have an enhanced visible-light absorption in comparison with graphene, which potentially can be important for photovoltaic applications.
The peak shift was found before for 
the same size of supercell in the case of parallel electric field at $1.11$ and $1.39$~eV for 
b-BNG and nonb-BNG-2~\cite{MORTAZAVI2019733}.
This shift in the transition has also been reported for a larger supercell of BN-codoped graphene with doped ratio of $25\%$ which corresponds to the same doping ratio as in our supercell~\cite{RANI201428}. 
It has been shown that the visible-light absorption vanishes at the low ratio of 
BN-doping~\cite{RANI201428}. 

Finally, the first peak indicating the $\pi \rightarrow \pi^*$ transition of nonb-BNG-1 is slightly shifted due to a small bandgap as is shown in \fig{fig05}(c).

\subsection{Thermal Properties}

In this section, electronic thermal conductivity, $k_e$, and specific heat, $c$, of the aforementioned four structures are presented in the temperature range of $T = 20-140$~K. 
Boltzmann equation implemented in the BoltzTraP software package is used to calculate both $k_e$ and $c$ near the charge neutrality point ~\cite{Madsen2006}.
In this temperate range, the electron and the lattice temperatures are decoupled for low dimensional graphene structures \cite{PhysRevB.87.035415,Gabor648} and
the rate of energy transferred between the charge carriers and the acoustic phonons is thus 
very small.
We therefore focus on the regime where the Fermi energy, $\mu \gg k_B T$
and $k_e \propto \sigma_e$, where $\sigma_e$ is the electrical conductivity. 

The investigation of $k_e$ defines how the Dirac charge carriers carry energy. This helps us to understand the efficiency of charge harvesting in graphene optoelectronic devices \cite{doi:10.1021/nl202318u}.
Electron thermal conductivity (a), electrical conductivity (b), and specific heat (c) are presented in \fig{fig06} for PG (brown), b-BNG (blue), nonb-BNG-1 (green), and nonb-BNG-2 (red) near the charge neutrality point. The electron thermal conductance increases proportionally to $T^2$ with temperature \cite{PhysRevB.76.115409}.
According to our calculations, the electronic thermal conductivity 
of b-BNG and nonb-BNG-2 is much larger that that of PG and nonb-BNG-1. This is attributed to the presence of a larger bandgap in b-BNG and nonb-BNG-2 in which hot
carrier–assisted charge transport and the electric conductivity is thus increased 
with temperature as is seen in \fig{fig06}(b).

The higher electronic thermal conductivity indicates that when
a charge carrier is excited by a thermal gradient, it can travel a larger
distance and excite additional carriers before it thermalizes with
the lattice.
Furthermore, a tunable electronic thermal conductivity implies a tunable
electronic specific heat (see \fig{fig06}c) which could be used to optimize 
bolometric applications of graphene. It thus may have applications in
thermoelectric technologies \cite{PhysRevX.3.041008, doi:10.1021/nl403967z, nano9050741, ABDULLAH2018223}.

\section{Conclusion}\label{Sec:Conclusion}

In this work, we have analyzed the electronic, stress-strain relations, the energy stability, the dynamic stability via phonon dispersion, the optical and the thermal properties of BN-bonded and non-bonded BC$_6$N structures. We have seen that the bandgap can be tuned by controlling the BN-bonds in the BC$_6$N structures depending on the electron delocalization processes via the aromacity of the structure. The presence of BN-bonds promotes the
BC$_6$N structure to a hard material which can withstand large strain. The anisotropy of the mechanical properties depend on the number of C-C, C-N, C-B, and B-N bonds in the structures. We demonstrate that energy stability, dynamic stability, and interaction energy 
depends on the BN-bonds and show that the graphene-like BC$_6$N with BN-bonds is the most stable among all studied atomic configuration structures. 
Furthermore, a high intensity peak in the dielectric function representing the absorption spectra in the visible range was observed in BC$_6$N structures with a large bandgap depending on the number of $\pi$-bonds. This may induce novel optical behavior in which BC$_6$N can be potentially used in optoelectronic devices. Finally, the electronic thermal conductivity and the specific heat show an outstanding performance of BC$_6$N with a large bandgap. Therefore, it can be used in thermoelectric devices.

\section{Acknowledgment}
This work was financially supported by the University of Sulaimani and 
the Research center of Komar University of Science and Technology. 
NRA would like to thank Swara Jalal for fruitful discussions.
The computations were performed on resources provided by the Division of Computational Nanoscience at the University of Sulaimani.  
 


\end{document}